\documentclass{llncs}
\usepackage[T1]{fontenc}
\usepackage[table]{xcolor} 
\usepackage{booktabs} 
\usepackage{caption}
\usepackage{xcolor} 
\usepackage{subcaption}
\usepackage{booktabs}
\usepackage{tabularx}
\usepackage{enumitem}
\usepackage{multirow}
\usepackage{graphicx}
\usepackage{xcolor}
\usepackage{wrapfig}
\usepackage[utf8]{inputenc}
\usepackage{tcolorbox}
\usepackage{soul}
\tcbuselibrary{skins}

\definecolor{Gray}{gray}{0.9}
\begin{document}
\title{Auditing M-LLMs for Privacy Risks: A Synthetic Benchmark and Evaluation Framework}
%
%
%
\author{
  Junhao Li\inst{1}\orcidID{0009-0008-1233-2236}
  \and
  Jiahao Chen\inst{2}\orcidID{0000-0002-5894-662X}
  \and
  Zhou Feng\inst{2}\orcidID{0009-0006-5301-7019}
  \and
  Chunyi Zhou\inst{2}\orcidID{0000-0003-0081-0946}\thanks{Corresponding author.}
}

\authorrunning{Li et al.}

\institute{
  Guangzhou University, China, People's Republic of\\
  \email{lijh@e.gzhu.edu.cn}
  \and
  Zhejiang University, China, People's Republic of\\
  \email{xaddwell@zju.edu.cn, zhou.feng@zju.edu.cn,zhouchunyi@zju.edu.cn}
}

\maketitle              
\begin{abstract}
Recent advances in multi-modal Large Language Models (M-LLMs) have demonstrated a powerful ability to synthesize implicit information from disparate sources, including images and text. These resourceful data from social media, also introduce a significant and underexplored privacy risk: the inference of sensitive personal attributes from seemingly daily media content. However, the lack of benchmarks and comprehensive evaluations of SOTA(state-of-the-art) M-LLM capabilities hinders the research of private attribute profiling on social media. Accordingly, we propose (1) \textbf{PRISM}, the first multi-modal, multi-dimensional and fine-grained synthesized dataset incorporating comprehensive privacy landscape and dynamic user history; (2) Efficient evaluation framework that measures the cross-modal privacy inference capabilities of advanced M-LLM. Specifically, PRISM is a large-scale synthetic benchmark designed to evaluate cross-modal privacy risks. Its key feature is 12 sensitive attribute labels across a diverse set of multi-modal profiles, which enables targeted privacy analysis. These profiles are generated via a sophisticated LLM agentic workflow, governed by a prior distribution to ensure they realistically mimic social media users. Additionally, we propose a Multi-Agent Inference Framework that leverages a pipeline of specialized LLMs to enhance evaluation capabilities. We evaluate the inference capabilities of six leading M-LLMs (Qwen, Gemini, GPT-4o, GLM, Doubao, and Grok) on PRISM. The comparison with human performance reveals that these MLLMs significantly outperform in accuracy and efficiency, highlighting the threat of potential privacy risks and the urgent need for robust defenses.
\end{abstract}

\section{Introduction}
Since the emergence of powerful multi-modal Large Language Models (M-LLMs), such as OpenAI’s GPT-4o~\cite{gpt4o2024} and Google’s Gemini~\cite{gemini2023}, models are no longer limited to analyzing text or images in isolation; they now exhibit sophisticated reasoning capabilities that allow them to synthesize clues across modalities \cite{wang2025capabilitiesgpt5multimodalmedical}. This leap in capability, particularly when applied to social media platforms, introduces a new privacy threat: M-LLMs could infer sensitive personal attributes by connecting the dots between a user's benign photos and textual posts.


To validate the real-world user perception of the privacy threat on social media, we conducted a large-scale \textbf{User Study} with 569 participants. As detailed in Section~\ref{motivation study} and illustrated in Figure~\ref{fig2}, our key findings highlight the problem's urgency:

\begin{itemize}
\item \textbf{User Concern:} 82.\% reported feeling ``very concerned'' or ``somewhat concerned'' about their information being used for privacy inference.
\item \textbf{Platform Trust:} Over 80\% believe that current privacy protections are ``insufficient'' or ``completely insufficient''.
\end{itemize}
Such findings provide strong evidence that multi-modal privacy risks constitute not merely a conceptual vulnerability but rather an imminent security challenge causing widespread user apprehension.

\begin{wrapfigure}{r}{0.6\columnwidth}
    \centering
    \includegraphics[width=0.6\columnwidth]{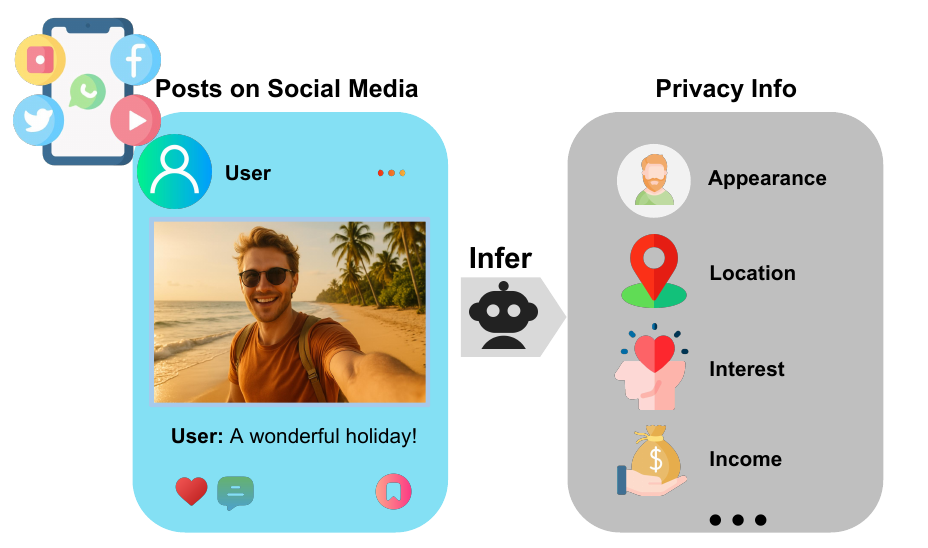}
    \caption{\textbf{Scenario Overview.} A photograph of an overseas vacation not only directly exposes the user’s appearance and location, but may also be inferred as personal attributes such as hobbies and income.}
    \label{fig:placeholder}
\end{wrapfigure}

As illustrated in Figure~\ref{fig:placeholder}, a single social media post can become a rich source for inference. While prior research has acknowledged attribute inference from single modalities like text~\cite{staab2024memorizationviolatingprivacyinference} or images~\cite{liu2025eyesherlockholmesuncovering,luo2025doxinglensrevealinglocationrelated}, these unimodal analyses fail to capture the integrated threat posed by M-LLMs. The privacy\cite{He2025AISPsurvey,10975146} risks arising from such cross-modal synthesis, where models can exploit a far richer context, remain largely unexplored.

Despite this well-founded user concern, a significant research gap hinders the systematic evaluation of this threat. \textbf{The academic community currently lacks a benchmark specifically designed for multi-modal privacy inference.} Specifically, existing datasets are often limited to single image-text pairs for general vision-language tasks (e.g., MS-COCO~\cite{cocodataset}) and do not contain the fine-grained, privacy-centric attribute annotations necessary for this analysis. 

\begin{figure}
\centering
\includegraphics[width=0.8\textwidth]{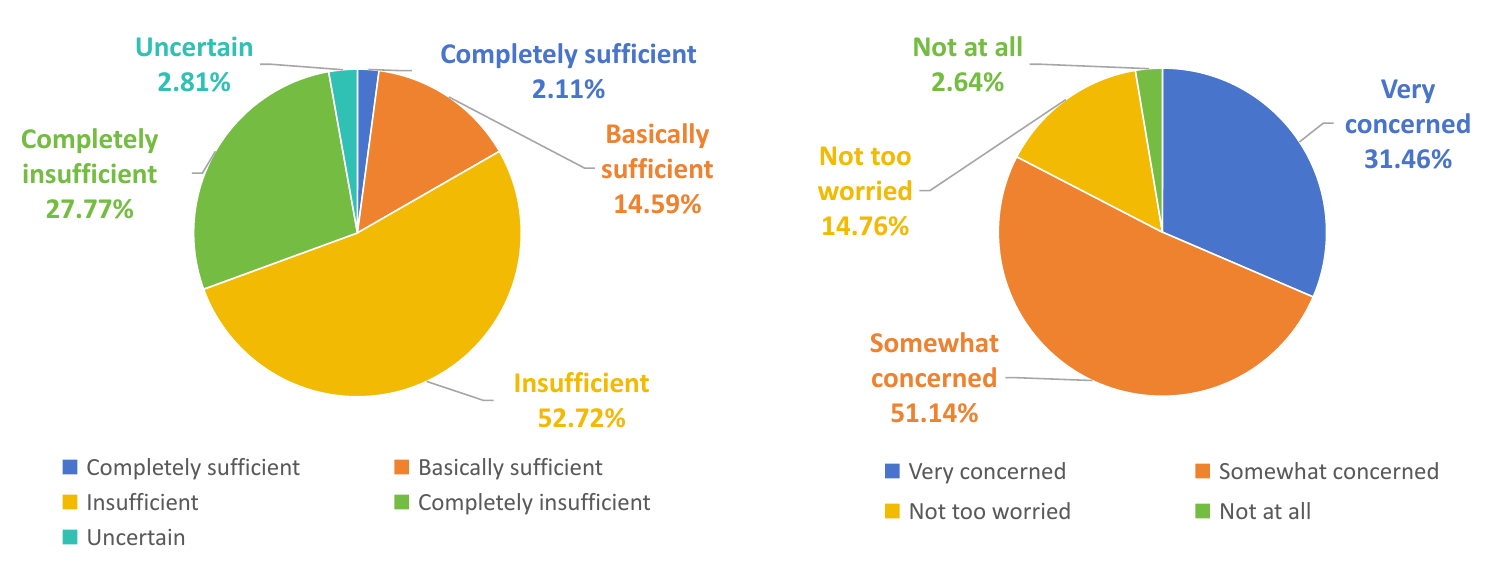}
\caption{Results from user study with 569 participants, highlighting public perception of multi-modal privacy risks. \textbf{Left:} User concern levels regarding the leakage of their personal information. \textbf{Right:} User perception of the adequacy of current platform privacy protections.
} \label{fig2}
\end{figure}

Furthermore, beyond the technical limitations of existing datasets, the direct use of real-world social media data for such research presents profound \textbf{ethical and legal challenges}. Essentially, this line of inquiry involves attempting to infer sensitive attributes that users have not explicitly consented to be analyzed for this purpose. Obtaining informed consent for a study designed to test the limits of privacy violation is a paradoxical and often insurmountable ethical barrier. Therefore, to conduct a systematic and reproducible investigation without compromising individuals' privacy, the creation of a synthetic benchmark is not merely a practical alternative, but a scientific and ethical imperative.

Besides, there are no established methodologies to probe how M-LLMs can be leveraged for multi-modal privacy inference, to quantify the privacy risks in the wild. This leaves a critical question unanswered: \ul{\textit{To what extent can M-LLMs systematically integrate disparate, multi-modal information from social media to construct complete and accurate user profiles?}}

To fill this critical technical and ethical gap, this paper introduces~\textbf{PRISM}, a novel framework and benchmark designed to systematically quantify the risks of cross-modal privacy inference on social media. Our framework leverages generative AI (e.g., Qwen, DeepSeek, Doubao) to create realistic synthetic profiles that mimic social media users, where ground-truth private attributes are linked to a series of posts containing subtle textual and visual clues. A comprehensive user study testified to the high quality of the generated dataset, with participants expressing strong satisfaction with both images and text. We then evaluate the capabilities of six leading M-LLMs (Qwen~\cite{qwen2_2024}, Gemini~\cite{gemini2023}, GPT-4o~\cite{gpt4o2024}, GLM~\cite{glm42024}, Doubao~\cite{doubao}, and Grok~\cite{grok2023}) on the PRISM benchmark for privacy inference. The privacy inference task is performed by our multi-agent framework, where specialized models first analyze clues from the text and image data, respectively. These unimodal findings are then channeled into a final multi-modal model that performs a holistic synthesis to produce the comprehensive inference. The results are assessed using a robust scoring system~\cite{wang2025mansounddemystifyingaudio} to ensure a fair and comprehensive analysis. Our major contributions are as follows:
\begin{enumerate}
\item To our knowledge, we are the first to systematically investigate the multi-modal privacy risk of M-LLMs profiling sensitive user attributes by synthesizing clues from disparate multi-modal social media content.
\item We develop an integrated framework for privacy risk assessment, consisting of PRISM, the first large-scale synthetic and multi-modal benchmark purpose-built for this task, and a Multi-Agent Inference Architecture to automate the evaluation and measure the potential risks in the wild.
\item We conduct a comprehensive evaluation of leading M-LLMs using PRISM. Supported by robust evaluation metrics and our novel architecture, our experiments systematically reveal the models' alarming proficiency and quantify the critical impact of multi-modality on privacy inference.
\item Through a direct comparison with human investigators, we provide the first quantitative evidence that M-LLMs can outperform them in both accuracy and efficiency for this task, highlighting the scalability of the threat.
\end{enumerate}

\section{Related Work}

\subsection{M-LLMs and Emergent Risks for Privacy Inference}
The emergence of M-LLMs~\cite{gemini2023} has dramatically enhanced AI's ability to interpret diverse data by integrating vision with language understanding~\cite{fang2025safemlrmdemystifyingsafetymulti-modal,gemini2023}. This introduces new privacy risks, as these models can infer sensitive information like geolocation~\cite{luo2025doxinglensrevealinglocationrelated} from seemingly benign content on social media. Unlike traditional attacks that require specialized skills, modern LLMs lower the barrier for non-experts and can achieve superhuman performance. For instance, a recent study showed that M-LLMs outperform humans in image geolocation with up to 21 times lower error distances, demonstrating their potential for large-scale, low-effort privacy attacks~\cite{haas2024pigeonpredictingimagegeolocations}. Concurrently, existing defense mechanisms like image perturbation~\cite{8237427} and prompt-based refusal often prove insufficient, highlighting an urgent need for privacy-aligned defense strategies.

\subsection{Attribute Inference from multi-modal Social Media Content}
The shift to M-LLMs has transformed social media analysis, enabling sophisticated reasoning from subtle multi-modal clues~\cite{info:doi/10.2196/59505} and introducing new privacy risks~\cite{DBLP:journals/eswa/NovikovaDK25}. These models can infer a wide range of sensitive attributes from routine posts~\cite{xu2025polarizedpatternslanguagetoxicity}, effectively democratizing the threat by lowering the barrier for non-expert adversaries~\cite{luo2025doxinglensrevealinglocationrelated}. The risk is particularly acute in the visual domain, where models like PIGEON~\cite{haas2024pigeonpredictingimagegeolocations} demonstrate superhuman capabilities in tasks such as image geolocation, sometimes achieving up to 21× lower error distances than humans~\cite{luo2025doxinglensrevealinglocationrelated}. This capability to turn subtle visual details into precise, sensitive data constitutes a tangible and scalable privacy attack vector.

Compounding this threat is the inadequacy of current defense mechanisms. Data-level strategies like image perturbation~\cite{Li_2019_ICCV,Feng2025Enkidu} often face a critical trade-off between effectiveness and content utility. Meanwhile, model-level safeguards~\cite{DBLP:conf/naacl/LuoCLWWFYL25}, such as refusing malicious prompts~\cite{ge2025llmsvulnerablemaliciousprompts}, are proving brittle and can often be circumvented by carefully engineered jailbreak prompts~\cite{298254}. This dangerous imbalance between rapidly advancing inference capabilities and lagging defense strategies highlights the critical need for standardized benchmarks to systematically evaluate and understand these privacy vulnerabilities.

\section{A Methodological Framework for Privacy Risk Evaluation}
\label{sec3}
Our methodology is structured to first establish the practical significance of multi-modal privacy threats through a user study, and then to detail the technical framework we developed to systematically evaluate these threats.
\begin{figure}
\centering
\includegraphics[width=0.75\textwidth]{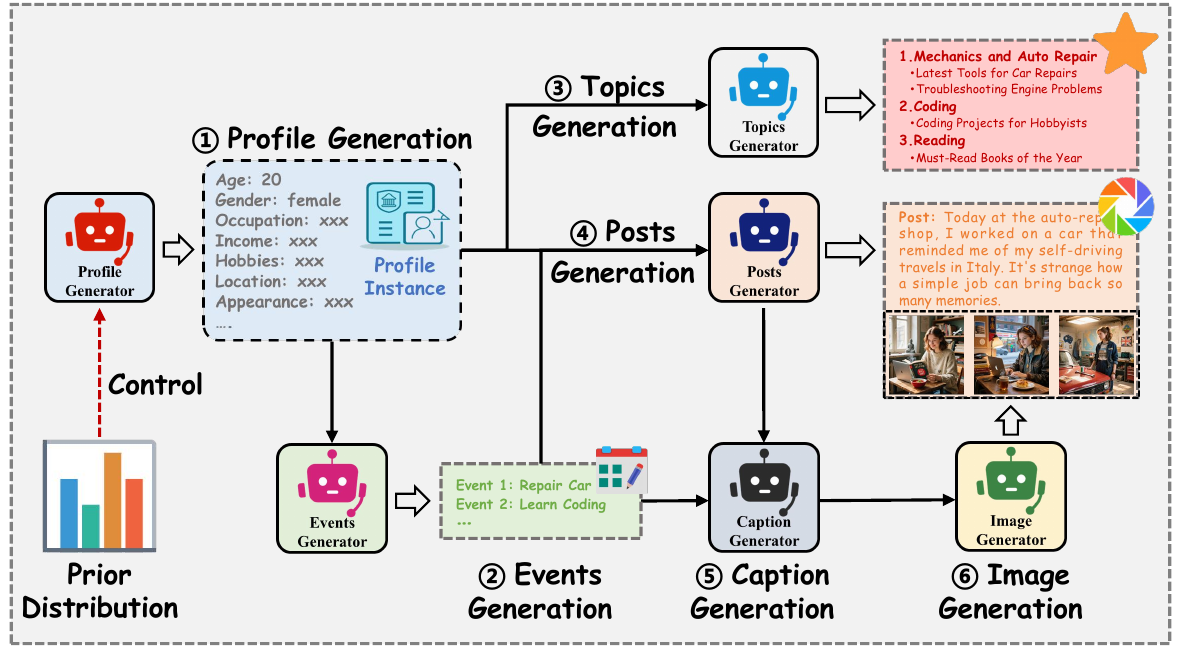}
\caption{The data generation workflow of PRISM begins by creating a realistic user Profile from a controllable Prior Distribution. Specialized generators then synthesize corresponding Events, Topics, Posts, and a Caption. Finally, an Image Generator renders a high-fidelity image from the caption, creating a complete and contextually consistent multi-modal post.
} \label{fig1}
\end{figure}
\subsection{Motivation Study}
\label{motivation study}
To validate the real-world significance of this threat, we first conducted a comprehensive user study. The survey was designed to create a holistic picture of the current privacy landscape, covering user demographics, social media~\cite{11036168} sharing habits, and awareness of AI-driven data analysis. We collected responses from 569 participants, representing a diverse cross-section of the online population. The respondents spanned a wide range of age groups and professional backgrounds, including students, IT professionals, educators, and individuals from various other industries, ensuring our findings reflect a broad user perspective.

The results, partially summarized in Figure~\ref{fig2}, reveal widespread public anxiety. A significant majority of respondents, 82.6\%, reported feeling ``very concerned'' or ``somewhat concerned'' about their publicly shared personal information being used for privacy inference or profit. Concurrently, over 80\% of users believe that the privacy protection measures currently offered by online platforms are ``somewhat insufficient'' or ``completely insufficient''. These findings strongly validate that privacy leakage stemming from multi-modal content is not merely a potential technical threat, but a pressing, real-world issue causing broad public apprehension.

Based on our analysis of multi-modal social media data, we define a comprehensive set of twelve sensitive attributes for evaluation: AGE, EDU, SEX, OCC, INC, REL, LOC, APP, SHP, HOB, HEA, and REG. The selection of these attributes was a deliberate process, informed by a comprehensive review of existing literature on privacy inference~\cite{wang2025mansounddemystifyingaudio,10184063} and the findings from our own user study. By synthesizing these sources, we consolidated the list to cover key facets of an individual's life, including demographics (AGE, SEX), socio-economic status (EDU, OCC, INC), personal life (REL, LOC, APP, SHP, HOB), and sensitive personal information (HEA, REG). This structured and well-grounded set of attributes provides the foundation for a holistic and meaningful evaluation of M-LLMs' inference capabilities.

\begin{tcolorbox}[colback=gray!15,colframe=gray!60!black ,title=Private Attributes of Adversaries' Interests,halign title=center,]
\begin{center}
\begin{tabular}{lll}
-- Age (AGE) & -- Occupation (OCC) & -- Hobbies (HOB) \\
-- Education (EDU) & -- Income (INC) & -- Location (LOC) \\
-- Sex (SEX) & -- Relationship (REL) & -- Appearance (APP) \\
-- Religious-Belief (REG) & -- Shopping-Habits (SHP) & -- Health-Status (HEA) \\
\end{tabular} 
\end{center}
\end{tcolorbox}

Our preliminary experiments on existing large-scale, open-source datasets, such as Reddit image-text pairs, showed that while M-LLMs could infer some private attributes, the results were highly inconsistent and difficult to evaluate systematically. We identified the critical bottleneck as the lack of a predefined, structured set of target privacy attributes, which caused the models' inferences to be unfocused, non-specific, and speculative. For instance, a model might output five attributes for one image but a completely different set of thirteen for another, including speculative categories. This absence of a consistent attribute schema made a fair and scalable evaluation impossible. This underscored the necessity for a purpose-built benchmark with a fixed set of privacy attributes, and therefore, to enable a rigorous and reproducible scientific investigation, we developed our own synthetic dataset and evaluation framework.

\subsection{A Synthetic Benchmark for Auditing Cross-Modal Privacy Risks}
Our evaluation is based on PRISM, a high-quality synthetic dataset generated through an automated agentic workflow that mirrors the complexity of real-world social media content. As given in Figure~\ref{fig1}, the generation process begins with a Profile Generator, an LLM that creates a ground-truth ``Profile Instance'' for each virtual user containing 12 sensitive attributes, which are sampled from a controlled prior distribution to ensure realistic diversity and representation.

\subsubsection{Prior Distribution} To ensure the generated dataset reflects the complexity and diversity of real-world user populations, the Profile Generator does not create attributes randomly. Instead, it samples from a carefully defined prior distribution grounded in empirical data to enforce realistic demographic constraints and inter-attribute correlations.

For instance, the location attribute is sampled according to the regional breakdown of global internet users, with higher probabilities assigned to regions with larger online populations, such as Asia, based on data from global digital reports~\cite{globaldigital2024}. The sex and age distributions are calibrated to align with demographic surveys of social media users . Crucially, we enforce logical correlations between attributes to prevent unrealistic profiles. The income level is conditioned on the user’s occupation, reflecting official wage statistics, and the selection of hobbies is conditioned on their age group, consistent with patterns found in sociological research . This principled, data-grounded approach is vital for creating a benchmark that poses a realistic challenge and yields meaningful insights.

Our workflow automatically generates the multi-modal dataset by synthesizing a rich user footprint through a sequence of five specialized generators, as illustrated in Figure~\ref{fig1}. The roles within this pipeline are assigned as follows:

\textbf{Events Generator: }Synthesizes a series of plausible life events based on the attributes in the ground-truth Profile Instance to provide narrative context.

\textbf{Topics Generator: }Enriches the user’s semantic profile by expanding on their hobbies and events to create a list of associated interests.

\textbf{Posts Generator: }Creates the textual content of a social media post, taking the user’s persona and a specific event as input to craft a natural language post.

\textbf{Caption Generator: }A critical intermediate step in our visual synthesis process. We observed in preliminary experiments that directly tasking a text-to-image model with interpreting abstract event and profile data yielded inconsistent and low-quality images. To solve this, the Caption Generator acts as a ``translator'', converting the event, topics, and post text into a detailed, descriptive prompt optimized for a text-to-image model. This ensures a high degree of control and relevance in the final generated image.

\textbf{Image Generator: }Takes the detailed caption from the previous step as input and renders a high-fidelity, contextually appropriate image to be paired with the text post, completing the multi-modal data point.

The final output of this pipeline is a complete, multi-modal dataset where each user profile is linked to a chronological series of posts. This structured approach ensures each data point is contextually consistent, providing a solid foundation for our subsequent evaluation experiments.

\subsection{The Multi-Agent Inference Architecture}
To perform the privacy inference simulation, we design a structured inference pipeline composed of three distinct nodes, each leveraging a specialized LLM.

\textbf{Textual Analysis Agent: }Acting as the textual specialist, this model is first responsible for processing all written content from a user's profile, such as posts and topics. It utilizes a foundational LLM to perform a deep analysis of this text, extracting a preliminary set of potential privacy-related clues in a structured format.

\textbf{Image Analysis Agent: }Operating in parallel, a visual specialist processes all the image data. For each of the user's images, this model prompts a vision-capable LLM with both the image and its corresponding text description. This method enables a context-aware analysis, allowing it to accurately identify objects, scenes, and other visual elements that could serve as privacy clues.

\textbf{Multi-modal Synthesis Agent: }Finally, the textual and visual clues extracted in the previous stages are channeled into the synthesis model, along with all the original data (including original posts, images, etc.). This model utilizes a powerful, vision-capable LLM to perform the final synthesis and inference. It integrates all the disparate pieces of information to generate a comprehensive and structured final user profile.

\section{Experiment}

This section presents a comprehensive evaluation of our proposed framework. We conducted two main experiments on  PRISM dataset. First, we evaluated six SOTA LLM on their privacy inference capabilities. Second, Human Evaluation demonstrates that these models significantly surpass human performance in both inference accuracy and efficiency, underscoring the scale of the automated threat. By systematically benchmarking multiple SOTA LLMs on PRISM and contrasting them with human baselines, we validate the robustness and representativeness of our benchmark and framework: the results reveal the universality of models’ inference capabilities across diverse attributes and highlight the severity of privacy risks in real-world contexts.

\subsection{Experimental Setup}
 All experiments were conducted on our PRISM dataset, comprising over 500 unique user profiles. Each profile is richly detailed, associated with 2-3 sets of images and corresponding posts, as well as more than 3 topics of interest with multiple associated articles. The ground-truth for each profile consists of 12 distinct privacy attributes as defined in Section~\ref{sec3}.
 
We evaluated six prominent LLMs: Qwen, GPT-4o, Gemini, Doubao, Grok, and GLM. The models were tasked with inferring user profiles using our Multi-Agent Inference Architecture. This architecture is composed of three distinct nodes: a text LLM that processes textual information (posts, topics, articles); an image LLM that analyzes visual data and its associated text; and a multi-modal LLM that aggregates the findings from the previous two nodes to generate the final, comprehensive inference result.

To ensure a fair and robust assessment, we employed an LLM-as-a-Judge protocol. The inference results from each target model were evaluated by two independent, powerful judge models: Qwen (qwen3-coder-480b-a35b-instruct) and GPT-4o. The judges scored the accuracy of each inferred attribute against the ground-truth on a scale of 0 to 100. The final score reported is the average of the scores from these two judges. We report results in the format of mean. To answer our research questions, we first conduct an automatic evaluation focused on the impact of multi-modality, followed by a human evaluation.

\subsection{Automatic Evaluation of Private Attribute Inference in M-LLMs}

To answer our research questions, we conducted an ablation study to quantify the impact of modality. 
\subsubsection{Ablation Study}The inference task was run under two conditions: (1) In the Text-Only setting, the inference pipeline was simplified to only use the Text LLM before final aggregation. (2) In the Multi-Modal setting, the full three-node architecture was employed, providing the models with both text and their corresponding images.

The comprehensive results are presented in Table~\ref{tab:average}. Our findings reveal that the inclusion of visual information can substantially enhance inference performance, though the degree of improvement varies across different models and attributes. The enhancement is particularly dramatic for attributes with strong visual correlations, such as sex and appearance. For instance, with the Qwen model, the inference accuracy for appearance surged from a mere 1.28 to 33.93 when introducing visual data. Similarly, for the sex attribute, several models achieved accuracy rates exceeding 95\% in the multi-modal setting, a stark contrast to the less reliable text-only results. This provides strong empirical support for our central hypothesis: visual data constitutes a rich, often essential, channel for privacy leakage that is effectively exploited by M-LLMs. 

\begin{table}[t]
\centering
\captionsetup{} 
\captionsetup[table]{skip=8pt} 
\caption{\textbf{Comprehensive performance comparison of six MLLMs on the PRISM benchmark.} The table details the inference accuracy, scored on a scale of 0 to 100, for each of the 12 privacy attributes under two distinct settings: Text-Only and Multi-Modal.}
\label{tab:average}
\begin{tabular*}{\textwidth}{l l @{\extracolsep{\fill}} rrrrrr}
\toprule[1pt] 
\textbf{Attribute} & \textbf{Modality} & \textbf{Qwen} & \textbf{GPT-4o} & \textbf{Gemini} & \textbf{Doubao} & \textbf{Grok} & \textbf{GLM} \\
\midrule 
\midrule
\multirow{2}{*}{AGE} & Text & 5.81 & 13.07 & 4.65 & 2.62 & 5.98 & 10.99 \\
& \textbf{Multi-Modal} & \textbf{58.83} & \textbf{15.56} & \textbf{65.98} & \textbf{23.14} & \textbf{57.64} & \textbf{42.84} \\
\hline
\multirow{2}{*}{EDU} & Text & 10.91 & 3.16 & 3.96 & 3.13 & 7.04 & 11.42 \\
& \textbf{Multi-Modal} & \textbf{38.33} & \textbf{3.87} & \textbf{7.20} & \textbf{5.43} & \textbf{2.67} & \textbf{24.14} \\
\hline
\multirow{2}{*}{SEX} & Text & 21.73 & 42.55 & 15.20 & 4.15 & 14.64 & 12.89 \\
& \textbf{Multi-Modal} & \textbf{98.90} & \textbf{77.26} & \textbf{99.82} & \textbf{88.58} & \textbf{96.39} & \textbf{96.55} \\
\hline
\multirow{2}{*}{OCC} & Text & 65.70 & 55.09 & 71.07 & 49.17 & 61.56 & 52.88 \\
& \textbf{Multi-Modal} & \textbf{71.84} & \textbf{67.25} & \textbf{78.39} & \textbf{67.66} & \textbf{71.78} & \textbf{76.50} \\
\hline
\multirow{2}{*}{INC} & Text & 2.35 & 0.28 & 0.42 & 0.12 & 0.19 & 0.67 \\
& \textbf{Multi-Modal} & \textbf{5.36} & \textbf{0.12} & \textbf{1.20} & \textbf{0.14} & \textbf{0.58} & \textbf{1.73} \\
\hline
\multirow{2}{*}{REL} & Text & 21.64 & 22.82 & 25.05 & 21.80 & 23.62 & 19.48 \\
& \textbf{Multi-Modal} & \textbf{28.71} & \textbf{24.70} & \textbf{27.29} & \textbf{22.43} & \textbf{26.71} & \textbf{10.78} \\
\hline
\multirow{2}{*}{LOC} & Text & 59.69 & 50.59 & 50.55 & 43.68 & 52.18 & 42.32 \\
& \textbf{Multi-Modal} & \textbf{48.98} & \textbf{47.15} & \textbf{56.36} & \textbf{47.42} & \textbf{50.15} & \textbf{37.78} \\
\hline
\multirow{2}{*}{APP} & Text & 1.28 & 12.96 & 9.54 & 0.28 & 0.67 & 1.50 \\
& \textbf{Multi-Modal} & \textbf{33.93} & \textbf{18.72} & \textbf{33.85} & \textbf{32.21} & \textbf{20.78} & \textbf{31.69} \\
\hline
\multirow{2}{*}{SHP} & Text & 35.11 & 27.98 & 41.64 & 25.73 & 35.35 & 20.24 \\
& \textbf{Multi-Modal} & \textbf{25.14} & \textbf{27.61} & \textbf{50.84} & \textbf{32.46} & \textbf{35.86} & \textbf{20.26} \\
\hline
\multirow{2}{*}{HOB} & Text & 70.90 & 72.23 & 68.80 & 68.69 & 78.98 & 61.63 \\
& \textbf{Multi-Modal} & \textbf{58.27} & \textbf{65.56} & \textbf{59.49} & \textbf{61.08} & \textbf{64.50} & \textbf{60.42} \\
\hline
\multirow{2}{*}{HEA} & Text & 40.52 & 37.55 & 38.75 & 28.02 & 40.63 & 24.68 \\
& \textbf{Multi-Modal} & \textbf{47.12} & \textbf{38.21} & \textbf{50.67} & \textbf{38.77} & \textbf{47.89} & \textbf{34.49} \\
\hline
\multirow{2}{*}{REG} & Text & 54.26 & 52.10 & 54.74 & 43.99 & 52.65 & 36.56 \\
& \textbf{Multi-Modal} & \textbf{52.48} & \textbf{54.26} & \textbf{59.09} & \textbf{56.91} & \textbf{52.53} & \textbf{44.83} \\
\bottomrule[1pt] 
\end{tabular*}
\end{table}
This ablation study confirms that the visual modality is not merely supplementary but often the primary channel for privacy inference in these systems.The experiment aims to answer two key research questions:

\textbf{RQ1: How proficient are SOTA models at inferring private attributes?}
Our findings indicate that current SOTA models possess a strong capability in profiling users from multi-modal data. As shown in the Multi-Modal results in Table~\ref{tab:average}, models like Gemini consistently lead in performance, achieving the highest scores across 7 of the 12 attributes. Certain attributes are exceptionally vulnerable. For instance, SEX is inferred with near-perfect accuracy by multiple models, with Gemini reaching a score of 99.82. Attributes that are often explicitly mentioned or strongly implied in social media discourse, such as OCC (Occupation) and HOB (Hobbies), also exhibit high inference scores, with top models like Gemini and Grok scoring 78.39 and 64.50, respectively. Conversely, attributes that are typically more private and less frequently disclosed, such as INC, prove to be the most difficult to infer, with all models scoring below 6. This demonstrates that while models are highly capable, their proficiency varies significantly depending on the nature of the private attribute.

\textbf{RQ2: What is the quantitative impact of visual information on this inference task?}
By comparing the Multi-Modal and Text-Only results within Table~\ref{tab:average}, we can quantify the dramatic impact of visual data. The inclusion of images serves as a powerful catalyst for privacy inference, particularly for attributes that are inherently visual.
The most striking example is APP. In the text-only setting, models struggled immensely, with an average score of just 4.37 across all models. However, when images were introduced, the average score surged to 28.53, an increase of over 550\%. For some models like Doubao, the score leaped from a mere 0.28 to 32.21. Similarly, the SEX attribute saw its average inference score skyrocket from 18.53 (Text-Only) to 92.92 (Multi-Modal), transforming the task from a difficult guess to a near-certainty. This is because images provide undeniable visual evidence of gender and appearance that text can only vaguely describe. The AGE attribute also saw a massive average performance gain, jumping from 7.19 to 43.99.

Interestingly, for a few attributes, the impact of visual data is less pronounced or even negative. For HOB, the average score slightly decreased from 70.22 in the text setting to 61.55 in the multi-modal setting. This suggests that for topics often explicitly stated in text, accompanying images may add little new information or even introduce ambiguity. A similar trend was observed for LOC with some models. This nuanced result underscores that the value of visual data is context-dependent, but for visually-grounded attributes, it represents the most significant vector for privacy leakage.

\subsection{Human Evaluation}
\subsubsection{Setup.} To further evaluate the effectiveness and efficiency of employing MLLMs to profile sensitive attributes from multi-modal content, we conducted a human evaluation. For this study, we recruited 20 graduate students, all majoring in computer science. Participants were familiar with and accustomed to using various search engines and LLM services. In this study, we randomly selected a subset of user profiles from our PRISM dataset. Participants were instructed to view the multi-modal posts (both images and text) associated with these individuals and infer the 12 privacy-related attributes. They were allowed to view the content multiple times as needed and permitted to use search engines or LLMs to retrieve relevant information, enabling more accurate inferences.

\begin{table*}[h]
\centering
\caption{Performance comparison between MLLMs and human participants. The table shows inference accuracy (\%) for each of the 12 attributes, the overall average (Avg), and the total time spent on the task in minutes.}
\label{tab:model_performance}
\resizebox{\textwidth}{!}{%
\begin{tabular}{@{\extracolsep{\fill}} l | r | *{12}{c} | c}
\toprule[1pt]
\textbf{Models} & \textbf{Spent Time} & \textbf{AGE} & \textbf{EDU} & \textbf{SEX} & \textbf{OCC} & \textbf{INC} & \textbf{REL} & \textbf{LOC} & \textbf{APP} & \textbf{SHP} & \textbf{HOB} & \textbf{HEA} & \textbf{REG} & \textbf{Avg} \\
\midrule
\midrule
Human   & 485.75 minutes& 21.64 & \textbf{35.00} & 100.00 & 55.00 & 0.00 & \textbf{30.00} & 56.25 & 38.12 & 12.50 & 16.83 & 50.00 & 30.00 & 37.11 \\
\hline
Doubao  & 75.6 minutes& 49.35 & 2.00 & 90.00 & 78.50 & 0.00 & 14.50 & 41.50 & 35.42 & 33.50 & 49.63 & 45.50 & 53.50 & 49.46 \\
Qwen    & 64.8 minutes& 46.83 & 18.50 & 80.00 & 79.00 & \textbf{2.00} & 23.00 & 47.00 & 23.01 & 27.50 & 54.54 & \textbf{53.50} & 44.50 & 49.95 \\
GLM     & 56.5 minutes& 50.42 & 14.00 & 98.00 & 74.50 & 0.50 & 11.00 & 39.42 & \textbf{38.88} & 15.00 & 48.70 & 37.50 & 41.50 & 47.45 \\
Gemini  & 196.3 minutes& \textbf{60.55} & 16.50 & \textbf{100.00} & \textbf{92.00} & 0.00 & 20.50 & \textbf{56.57} & 33.00 & \textbf{50.00} & \textbf{56.99} & 50.00 & \textbf{54.50} & \textbf{54.25} \\
\bottomrule[1pt]
\end{tabular}%
}
\end{table*}

Comparison between Humans and MLLMs. The results of our Human Evaluation study are detailed in Table~\ref{tab:model_performance}. The findings reveal a significant performance gap between MLLM agents\cite{gan2024navigatingriskssurveysecurity} and human participants across both accuracy and efficiency. In terms of average inference accuracy, all evaluated models surpassed the human average of 36.28. Notably, our top-performing model, Gemini, achieved an average accuracy of 54.25, representing a substantial absolute improvement of nearly 18 points over human performance. This superiority was particularly evident in attributes requiring deep contextual understanding or broad knowledge, such as OCC, where Gemini scored 92.00 compared to the human score of 55.00.

Furthermore, the M-LLM agents demonstrated a staggering advantage in time efficiency. Human participants required a total of 485.75 minutes to complete the task, whereas the fastest model, GLM, finished in just 56.5 minutes—over 8.5 times faster. This highlights the most critical aspect of this threat: its scalability. M-LLMs can perform automated, large-scale privacy reconnaissance at a speed and scale that is simply unattainable for human actors, posing a severe and scalable risk to user privacy. We found a positive correlation between human and LLM inference capabilities, suggesting that attributes easier for humans to infer are generally also easier for the models to infer.


\section{Discussion and Potential Defenses}
Our findings underscore the urgent need for effective defenses against cross-modal privacy inference. Current research has explored two main categories of defense, each with a significant trade-off. The first, data-level defense, involves perturbing inputs with methods like Gaussian noise for images \cite{luo2025doxinglensrevealinglocationrelated} or signal jamming for audio \cite{wang2025mansounddemystifyingaudio}. While this approach can directly disrupt clues, it faces a critical dilemma: strong perturbations severely degrade content quality and user experience, whereas weaker ones are often insufficient to fool sophisticated model reasoning. The second category, model-level defense, uses prompt engineering to compel models to refuse sensitive requests \cite{luo2025doxinglensrevealinglocationrelated}. The advantage is that it preserves the original data, but the drawback is that models often become over-defensive, rejecting benign queries and thus harming their overall utility.

Given these limitations, we argue that future defense strategies must be more intelligent and multi-layered. At the user level, social media platforms could integrate proactive warning systems that analyze content before a user posts, alerting them to potential privacy risks and empowering them with informed control. At the model level, more advanced techniques like In-context Unlearning \cite{pawelczyk2024incontextunlearninglanguagemodels} offer a promising direction. This method can dynamically suppress a model's ability to associate specific features with sensitive attributes on a per-query basis, offering a targeted defense without permanently degrading the model's capabilities. We believe that combining proactive user-side warnings with intelligent model-side suppression is key to building a robust defense ecosystem for the future.

\section{Conclusion}
In this paper, we systematically investigated the emergent threat of cross-modal privacy inference on social media. Our work makes two primary contributions. First, we introduce PRISM, a novel Dataset Benchmark that is multi-modal, multi-dimensional, fine-grained, and designed for quantifiable privacy risk analysis. Second, using this benchmark, we established the first comprehensive Model Performance Benchmark for this task by evaluating six leading M-LLMs.Our findings reveal their alarming proficiency, with the inclusion of visual data boosting inference accuracy by over 550\% for key attributes like appearance. Furthermore, the models significantly surpassed human investigators in both accuracy and efficiency. They completed the task over 8.5 times faster, confirming the scalability of the threat. This work validates the feasibility of automated privacy attacks and provides a foundational framework to facilitate future research into robust defenses.
\begin{credits}
\subsubsection{\ackname}
This work was partly supported by the National Key Research and Development Program of China (No.~2022YFB3102100), 
the NSFC–Yeqisun Science Foundation (No.~U244120033), 
the National Natural Science Foundation of China (Nos.~U24A20336, 62172243, 62502432, 62402425, 62402418), 
the China Postdoctoral Science Foundation (No.~2024M762829), 
the Zhejiang Provincial Natural Science Foundation (No.~LD24F020002), 
and the Zhejiang Provincial Priority-Funded Postdoctoral Research Project (No.~ZJ2024001).
\end{credits}

\bibliographystyle{splncs04}
\bibliography{references}
\end{document}